\let\accentvec\vec
\documentclass[iop]{emulateapj}

\let\vec\accentvec
\usepackage{amsmath}
\usepackage{amssymb}
\usepackage{graphicx}
\usepackage{hyperref}
\usepackage{tocbibind}

\begin{document}
\newcommand{\beq}{\begin{equation}}
\newcommand{\eeq}{\end{equation}}
\newcommand{\ben}{\begin{enumerate}}
\newcommand{\een}{\end{enumerate}}
\newcommand{\bit}{\begin{itemize}}
\newcommand{\eit}{\end{itemize}}
\newcommand{\barr}{\begin{array}}
\newcommand{\earr}{\end{array}}
\newcommand{\D}{\partial}
\newcommand{\DD}{\frac}
\long\def\symbolfootnote[#1]#2{\begingroup%
\def\thefootnote{\fnsymbol{footnote}}\footnote[#1]{#2}\endgroup}

\title{On the viability of gravitational Bose-Einstein condensates as alternatives to supermassive black holes }

\author{Hujeirat, A.A.\altaffilmark{1}}

\altaffiltext{1}{ IWR - Interdisciplinary Center for Scientific Computing, University of Heidelberg, INF 368, 69120 Heidelberg, Germany}

\begin{abstract}

Black holes are inevitable mathematical outcome of spacetime-energy coupling in general relativity.
Currently these objects are of vital importance for understanding numerous phenomena in astrophysics and cosmology.

However, neither theory nor observations have been capable of unequivocally prove the existence of black holes or
granting us an insight of what their internal structures could look like, therefore leaving researchers to speculate
about their nature.\\

In this paper the reliability of supermassive Bose-Einstein condensates (henceforth SMBECs) as alternative to
supermassive black holes is examined. Such condensates are found to suffer of a causality problem that terminates their
cosmological  growth toward acquiring masses typical for quasars and  triggers their self-collapse into  supermassive black holes (SMBHs).

It is argued  that a SMBEC-core most likely would be subject to an extensive deceleration of its rotational frequency
as well as to vortex-dissipation induced by the magnetic fields that thread the crust, hence diminishing the superfluidity
of the core.
Given that rotating superfluids between two concentric spheres have been verified to be dynamically unstable to non axi-symmetric
perturbations,
we conclude that the remnant energy stored in the core would be sufficiently large to turn the flow turbulent and
dissipative and subsequently lead to core collapse bosonova.

The feasibility of a conversion mechanism of normal matter into bosonic condensates under normal astrophysical
conditions is discussed as well.

We finally conclude that in lack of a profound theory for quantum gravity, BHs should  continue to be the favorite proposal
for BH candidates.\\ \\

\end{abstract}


\section{Introduction}

Before 227 years ago and 118 after Newton's gravitation (1666), John Michell (1787) was the first to anticipate the existence of
a critical radius, $R_H$, below which even the light cannot escape the gravitational pull of the central point mass object.
Based on Newtonian gravity, he equated the potential energy $E_p$ of the object of mass M to the kinetic energy $E_K$
and obtained: $R_{crit} = 2GM/c^2,$ where "c, G" are the speed of light and the gravitational constant, though
the speed of light was still uncertain and the possible existence of such light-capturing objects in nature was merely a
fictitious proposal.

In 1915 presented Einstein his profound theory of gravitation, therein postulating the energy as a field in
four-dimensional spacetime and posting it as source for curvature. Mathematically, Einstein field equations read:
\beq
\rm G_{\mu\nu} = \kappa T_{\mu\nu},
\label{EFE}
\eeq
where $\rm G_{\mu\nu}, T_{\mu\nu}$ are the 2-rank Einsteins's geometrical and energy-momentum tensors, respectively. The coefficient
$\rm \kappa = 8\pi G/c^2 \approx 1.863\times 10^{-27}.$ \\
Obviously, unless there is a unusually strong accumulation of energy in a relatively small volume, the LHS and the RHS of Eq. (\ref{EFE})
can be conveniently decoupled,  therefore justifying the Newtonian theory in the weak field limit in flat spacetime.

Shortly thereafter the GR was published in (1915), Karl Schwarzschild presented a solution for
the field equations that corresponds to a spherically symmetric object of mass M embedded in vacuum. This solution
was the first theoretical proof that BHs are inevitable products of general relativity (GR), where the spacetime becomes indefinitely warped. \\

Several years later, Subrahmanyan Chandrasekar (1931) proposed a framework for the BH-creation. Accordingly, sufficiently massive
stars could in principle collapse under their own self-gravity, as neither the thermal nor the degenerate pressures could counter-balance
the enormous gravitational attraction. \\

From the point of view of Schwarzschild's solution, a massive star could contract in a quasi-stationary manner to
finally undergo a dynamical collapse into indefinitely small size at the center. However, a sufficiently distant observer
would be able to observe the collapsing matter up to the horizon, but not beyond.\\

In the early times, this scenario was rejected both by theorists and observers alike and in particular by Einstein and Eddington. Only at
the beginning of the sixties the BH-proposal was revived, when BHs appeared to be the only reasonable objects
to explain the origin of the vast energy output observed to liberate from quasars.

Since then the BH-proposal has been repeatedly adopted and their mass-regime has been continuously expanded
to finally span the whole mass spectrum, ranging from the microgram scale up to billion solar masses.\\

Although BHs are unobservables by definition, the dynamical behavior of matter and stars in their vicinity
in principle should disclose the properties of these objects. In particular, the distinguished depth of their gravitational well,
enables BHs to convert a significant portion of the potential energy of
matter or objects approaching the BH into other forms of energy,  such as thermal or magnetic energy.
In fact there are additional observational techniques that are used for predicting the depth of the gravitational well of BHs,
e.g., the iron $K~\alpha$ emission lines associated with rotating plasmas in accretion disks, the Lorentz factor of
 ejected plasmas from their vicinity and possibly through gravitational waves detectors to be employed in the near future
\citep[see][for additional detection methods]{mueller2007}.\\

Most BHs in binaries are considered to have masses of the order of $\rm 10\, M_\odot$.  The massive ones are generally found at the center of galaxies
with masses ranges between several million up to several billions solar masses. In most cases they are observed to be radiatively active, implying
therefore that they are accreting from or ejecting matter into their surrounding.

Nevertheless, these techniques hint to the existence of objects that differ from normal or compact stars, such as old stars,
white dwarfs and neutron stars; however they definitely are unable to determine their nature.\\ \\

\noindent{\bf Difficulties with the classical BH-Proposal}\\\\
Consider an object of Mass M and spin=0 embedded in vacuum (see Fig.\ref{Fig1}), how does the spacetime around this object
look like?\\
 From the point of view of GR, the equations to be solved are $\rm G_{\mu\nu}=0.$\\
The solution to this problem is a metric, which was forwarded to Einstein by Schwarzschild just two months
after the former published his theory of GR. The metric reads:
\beq
\rm ds^2 = c^2(1-{r_s}/{r})dt^2 - \DD{dr^2}{1-{r_s}/{r}} - d\bar{\Omega}^2,
\label{SchrwarzschildEq}
\eeq

where $\rm ds,~c,~t,~ r_s$ correspond to the proper distance between two events in this spacetime, the
speed of light, the time measured by a fixed observer at infinity and the Schwarzschild radius $r_s = 2GM/c^2,$
respectively.
The term $\rm d\bar{\Omega}^2 = r^2 d\Omega^2 = r^2(d\theta + sin^2\theta~d\varphi^2)$ correspond to a
differential area on the surface of a sphere of radius r. \\

As it will be shown below, the Schwartzschild's proposal raises various fundamental questions in GR
rather than providing a solution to a problem:

\begin{figure}
\centering {\hspace*{-0.2cm}
\includegraphics*[width=8.0cm, bb=0 268 567 540,clip]{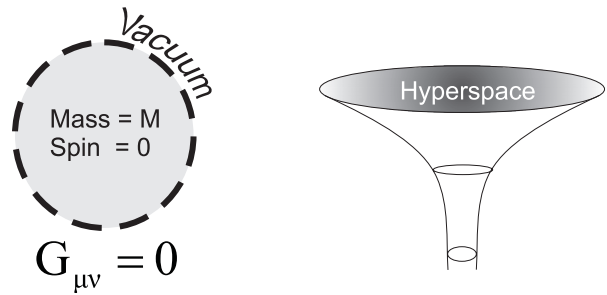}}
\caption{\small  The gravitational field of a non-rotating object of mass M embedded in vacuum
                 is equivalent to a curved spacetime having the event horizon as a boundary.
                  The hyperspace on the right is a projection of the four-dimensional curved spacetime.} \label{Fig1}
\end{figure}
\begin{itemize}
  \item
The Schwarzschild solution contains two singularities: r=0 and $\rm r=r_s.$ While the former is intrinsic and unremovable, the
latter one can be removed by appropriate coordinate transformation, e.g. using the Eddington-Finkelstein coordinates
 \citep{Hobsonbook}. The contraction of the Riemann curvature tensor at the event horizon:
 $R_{\mu\nu\sigma\tau}R^{\mu\nu\sigma\tau}$ yields a well-defined finite number, which implies that the curvature
 at this radius is well-behaved and that a co-moving observer will feel nothing special as she/he crosses the even horizon.\\
 In fact it was shown by \cite{Penrose1965} and \cite{Hawking1967} that if Einstein general theory of relativity is correct and
 the stress-energy tensor satisfies certain positive-definite inqualties, then spacetime singularities are inevitable.
 This may imply that all quantum information generally associated with the matter-building the BH will be completely destroyed.

 Moreover, as the lightcones tend to close up near the event horizon, $r=r_s,$ our connection
 to observers approaching the event horizon will be continuously weaker and will be completely lost inside $r=r_s,$ which
 implies that we will never know if this observer ever crossed the horizon.\\

 This raises a serious question about the progenitor of BHs. If these were massive stars that run out of energy generation
 via nuclear fusion at their centers followed by a dynamical self-collapse under their own self-gravity,
 then from our point of view as distant observers they must be still collapsing and would cross the event horizon with the speed of light
 after infinite time.
 Equivalently, the progenitors of BHs are massive stars that collapsed into rings of matter that surround the event horizon and whose particles approaching the horizon, but which will never reach or cross it.

\item  A photon of energy $E_0 = h\nu_0$ measured in our frame will be infinitely blue-shifted at the event horizon. Assuming global
energy conservation, this
 indefinite gain of energy must be extracted from the field and therefore would cause a non-negligible perturbation to the spacetime.
 As a consequence, each photon that approaches the horizon will cause a measurable change of spacetime curvature around the
 BH, which mathematically means, that the solution may not be stable against external perturbations.

 \item The entropy problem of BHs has been investigated by  \cite{Bekenstein1973}, who found that the entropy of a BH
 scales as the number of Planck spheres accommodate-able within the 2D projection area of a BH, or equivalently:
 \beq
 S \sim k_B (\DD{r_S}{\ell_P})^2 \approx k_B \times 10^{77}~(\DD{M}{M_\odot})^2.
 \label{EntropyEq}
 \eeq
$\rm k_B, and~ \ell_P$ correspond to  the Boltzmann constant and the Planck length, respectively. \\
Assuming a true association of the thermodynamical variables with the BH thermodynamics, then the entropy of progenitor
appears to increase by at least twenty orders of magnitude during a transition into a BH phase. This may imply that a large amount of
information is hidden behind the horizon  and that a deep holographical connection between the horizon as a
surface and to the BH as bulk is at work \citep{Padmanabhan2006}

\item The information paradox repeatedly discussed by theoreticians has not been satisfactorily solved yet.

Noting that the phenomena of particle-wave duality in quantum theory would enable particles to cross the event horizon as waves,
then the information about the pre-collapsed state of matter must be still found in the non-vanishing wave-tail. Such extraction
of information is possible due to the unitarity requirement of the Hamiltonian operator describing the quantum state of the
infalling particles. However, the only retrievable informations from BHs is Hawking-quanta, which is of black-body type and therefore featureless (s. Fig. \ref{Fig2}).

 Similar to the emitted black body radiation from
the sun's photosphere, these photons transmit highly diffused information about their paths inside the Sun due to their random-walk motions
in optically thick media.
Thus the Hawking radiation carry information about the physical conditions at or outside the interface,
where they make abrupt transition from the opaque spacetime domain inside the event horizon
into the optically thin and roughly flat spacetime outside it \citep[s.][for further details]{Preskill1992}.

 Also, it is not clear, whether energy stats of matter and the behavior of atoms in the vicinity of the event horizon would obey
 a Maxwell-Boltzmann like-distribution and whether the Planck function would continue to properly describe
 the photon spectrum.

\item The Schwarzschild solution (Eq. \ref{SchrwarzschildEq}) unequivocally shows, that GR is capable of singling out two events in
the spacetime and declare them to be singular in a deterministic manner. However, this approach is in complete contradiction to the
fundamental principles of quantum field theory and in particular to the Heisenberg uncertainty principle (Fig. \ref{Fig2}).

\item While the progenitors of stellar black hole are well-understood, the evolution of supermassive black holes (SMBHs)
is still a controversial issue. It is believed that the Pop III stars in the early universe must have been sufficiently massive, as
 the corresponding Jeans mass was of order $10^3 M_\odot$ \citep{Bromm1999}.

 Such metal-free massive stars must have formed at cosmological redshifts between $  8 \leq z \leq 20 $ and should have
 collapsed relatively fast to form the first massive black holes.  Since then their mass must have grown
 exponentially through repeated mergers and accretion of matter to acquire  typical quasar masses of the order $10^9 M_\odot.$
  However, the cosmological simulations were able to show certain condensations of clouds but not the final mass that
  exclusively go to form the massive star. In fact such modeling is associated with various numerical difficulties
  that severely limit the reliability of such large scale simulations.\\

 Furthermore,  the mass spectrum of BHs suffers of a gab which ranges from 100 to 10000 $M_\odot.$ The origin of this gab is
 unclear,  especially because  this range of masses could have been conveniently
  covered by the collapse of Pop III type stars.
  \end{itemize}

\begin{figure}
\centering {\hspace*{-0.2cm}
\includegraphics*[width=2.50cm, bb=71 240 255 448,clip]{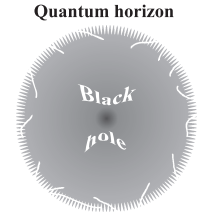}}
\caption{\small  The modern picture of a quantum horizon governed by quantum fluctuations and
quantum tunneling, giving rise  to the emission of quanta. The semi-classical approach of Hawking predicted
that
a BH of mass M emits BB-radiation at temperature $T_{BB} = 1/(8\pi G M).$} \label{Fig2}
\end{figure}


\section{\bf The long way toward BH-alternatives}

During the last decade several alternatives to BHs have been proposed. A considerable attention
was given  to dark energy stars (henceforth DESs) and gravitational vacuum stars
(henceforth Gravastars)  that have been proposed by \citet{Chapline2001, Mazur2004}.\\

Inspired from the $\Lambda-$cold dark matter cosmology \citep[$\Lambda CDM$; ][]{Peacock2001}), in which the vacuum energy is considered to
be responsible for the accelerating expansion of the universe, one may replace the inner-horizon region of a BH by
a vacuum-like core or a De Sitter spacetime. \\
In this case  the effect of gravity is a repulsive rather than impulsive.
As in the case of dark energy in cosmology, the energy density in the core is set to be constant whereas
the pressure is equal to the negative energy, i.e., $P=-\rho$.\\
Such a negative pressure is a typical phenomena on the scale of quantum fluctuations as the Casimir-effect
shows.
Furthermore, the  equation of state (EoS), $P=-\rho$, is an inevitable consequence of Einstein's field equations
describing an isotropic and homogeneous universe. Using the Robertson-Walker metric,
 the Friedmann equations  yield the following evolutionary equation for the scaling factor, "a":
\beq
           \DD{\ddot{a}}{a} = -\DD{4\pi G}{3}(\rho + 3p) + \DD{\Lambda}{3},
           \label{CosmoAccelerate}
\eeq
where $\Lambda$ denote the cosmological constant and $\ddot{a}$ corresponds to the second time derivative of "a"
\citep[see][for further details]{Peacock2001}.\\
As revealed by cosmological observations, including WMAP and high-redshift Type Ia supernovae \citep{WMAP2011},
 $\ddot{a}>0,$
which implies that the RHS of Eq. (\ref{CosmoAccelerate}) must be positive. However, if vacuum energy density is
due to zero-point energy of  quantized fields, then $\Lambda$ must be a Lorentz invariant and therefore can be
traced back in the history to make it negligibly small compared to the other terms in the equation. Consequently,
the term $\rho + 3P$ must be negative, hence $P< -\rho/3$, or generally, $P= \omega\rho,$ where  $\omega< -1/3$.
 In fact, recent WMAP cosmological observations reveal a very narrow range for
 $\omega:$  $-1.24 \leq \omega \leq -0.86$ \citep{WMAP2011},
 implying therefore that $P=-\rho$ as EoS for the vacuum core can be safely used.\\

\begin{figure}
\centering {\hspace*{-0.2cm}
\includegraphics*[angle=-0, width=4.50cm, bb=0 126 348 294,clip]{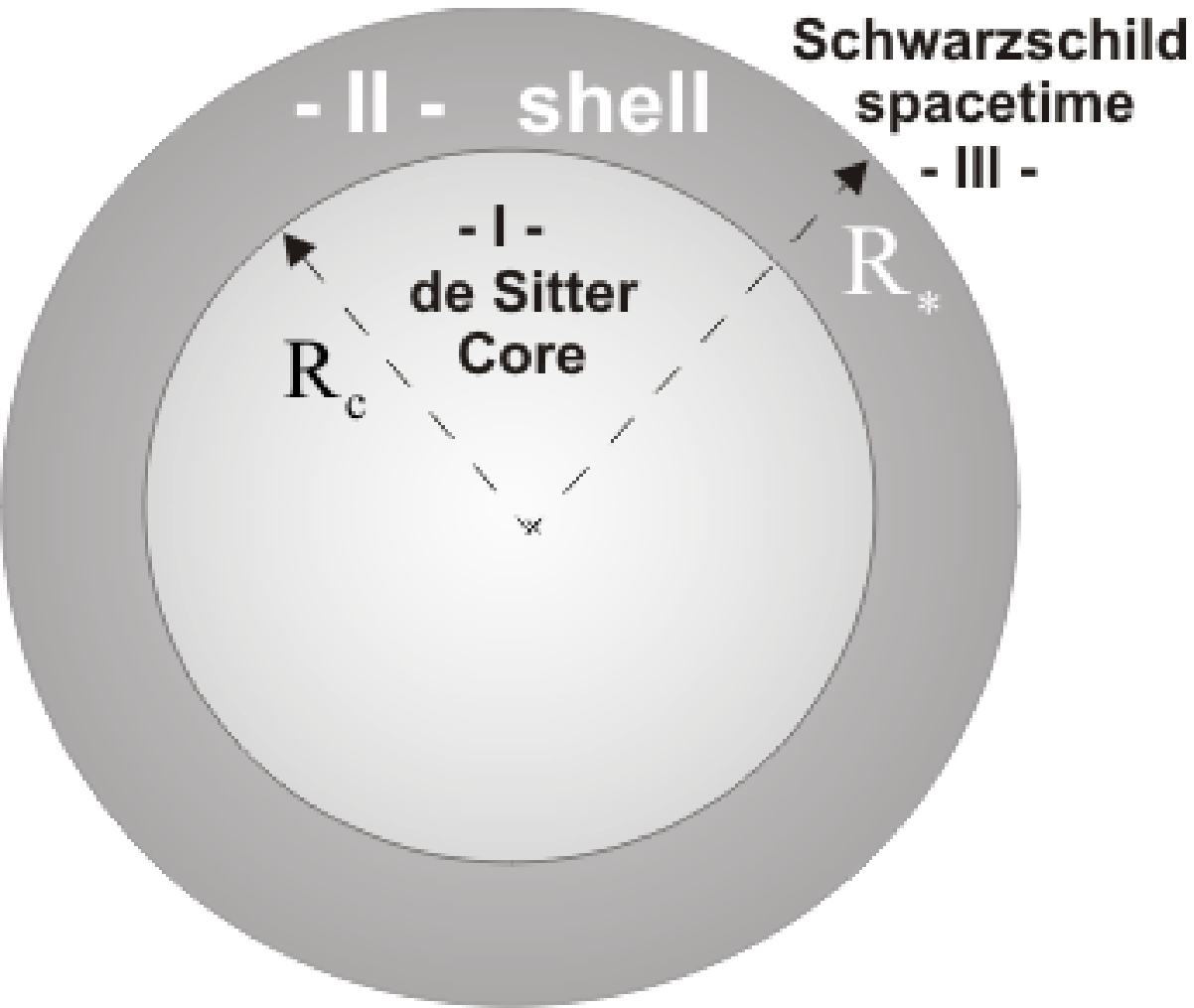}}
\caption{\small The gravastar model consists of three sub-domains: a central de Sitter core, a Schwarzschild
empty space outside the object and  a thin spherical shell in-between filled with normal matter. }  \label{Fig3}
\end{figure}

In the case of a Gravastar (\cite[see e.g.][]{Mazur2004}), the object consists of the following main three  domains (s. Fig. \ref{Fig3}):
\begin{enumerate}
  \item Region I: $0\leq r\leq R_C,$ where $R_C$ is the core radius. This region is governed by de Sitter spacetime governed by the
    EoS   $P=-\rho.$
  \item Region III:  $r>  R_\star,$ where $R_\star$ is the radius of the object. In this domain  $\rm T^{(matter)}_{\mu\nu}$ and $\Lambda $
  are set to vanish completely, hence the EoS reads: $P=\rho=0.$
  \item Region II: $R_C < r <R_\star$  corresponds to a spherical shell filled with normal matter that obeys the EoS: $P=\rho.$
\end{enumerate}
The aim is then to search for a global solution to the field equations:
\beq
    G_{\mu\nu} =  \kappa G_{\mu\nu} + g_{\mu\nu} \Lambda,
\label{MMMetric}
\eeq
that fulfills the boundary conditions across the different interfaces separating the three regions.
The solution procedure relies on proposing the following metric solution:
\beq
      \rm ds^2 = f(r)dt^2 - \DD{dr^2}{h(r)} - d\bar{\Omega}^2,
\label{MMMetric}
\eeq
where $f(r),\text{ and }h(r)$ are metric coefficients to be found.
Unlike Newtonian physics in Eucliedian geometry, the distribution of energy in each region
may have a significant impact on the curvature of spacetime. Hence a GR-consistent matching procedure,
such as the Israel junction condition \cite{Israel1966} is required in order to construct a global
solution that satisfies the conditions in the three different domains.\\

One possible solution runs as follows:
\begin{itemize}
  \item In the core region,   $f(r) = C\,h(r) = C\,(1-H^2_0 r^2),$ which is similar to the metric coefficients in
   the case of de Sitter spacetime in cosmology.  The coefficient $H_0$ is an integration constant that is
   analogous to the cosmological Hubble constant, though it has completely different value.
  \item In the outermost region, the solution is identical to that of Schwarzschild, i.e.,\\
  \[
   f(r) = h(r) =  1 - \DD{r}{r_s}.
  \]
  \item In the intermediate region and in the limit $R_C \longrightarrow R_\star,$ the following analytical solution
  was obtained:
 \[
   h(r) =  1 - \DD{2Gm(r)}{r\,c^2} \simeq \epsilon,
  \]
  where $m(r)$ is a continuous mass function, so that $dm(r) = 4\pi \rho r^2 dr.$ $\epsilon$ is an integration constant of order the Planck mass $M_{P}$ divided by the
  mass of the object. Most importantly, the integration constant  $\epsilon$ is strictly positive, which implies that $r_s$ is less than $r_\star,$ therefore prohibiting the formation of an event horizon.\\
  Following \cite{Mazur2004}, the thickness of the shell was estimated to be of order: $ \ell \sim \surd{\ell_P  r_s}\approx 10^{-14}\,$cm,
   where $\ell_P$ is the Planck length. Thus, for a one solar mass gravastar , $\ell$ would hardly exceed the radius of a proton.
\end{itemize}
Thus the shell is actually an extremely thin membrane rather than a normal matter-contained shell. Moreover,
the model requires a conversion mechanism  of unknown nature to operate efficiently at the base of the membrane,
whose function is to enforce normal matter to undergo a phase transition into a de Sitter vacuum state.\\ \\

\noindent{\bf The reliability of the gravitational Bose-Einstein condensates as BH-alternatives }\\

Similar to cosmology,   vacuum-dominated cores must expand. The effect of the membrane
is then to limit/decelerate the expansion rate, so to maintain these object in hydrostatic equilibrium.
\citep{Ghezzi2011} showed that an anisotropic pressure is required for ensuring dynamical stability of
dark energy objects, though a physical origin was not provided.\\

On the other hand, all normal astrophysical objects known to date, including the relativistic ones,
are found to have compactness parameters that are strictly less than $1/2$ (s. Table \ref{Table1}).\\
\citet[][see also the references therein]{Visser2003} found that under a variety of stellar-structure conditions,
 the compactness parameter has an  upper limit: $ \mathcal{C}< 4/9.$ Therefore, a one solar mass
 dark energy object (DEO)  with a radius $R_\star = r_s + \ell \approx (1+ \sqrt{\epsilon})\,r_s$ would have the compactness parameter
  $\mathcal{C} = 0.5 - 10^{-14},$  which implies that DEOs, if isolated, are almost indistinguishable from
  their BH-counterparts.\\
However, the extraordinary compactness of these objects, while having solid surfaces, rises several fundamental
questions about their reliability and viability in nature as elaborated in the following. \\ \\

\begin{table}[htb]
\begin{centering}
\begin{tabular}{l|c}
   & $\mathcal{C} = GM_\star/r_\star c^2$\\
  \hline
  Sun & $< ~10^{-5}$ \\
  White dwarfs & $<~ 0.0004$ \\
  Neutron stars & $<~ 0.16$ \\
  Quark star${}^*$ &  0.37 \\
  Dark energy star${}^*$ & $\approx \DD{1}{2}(1 - \sqrt{\epsilon})$ \\
  Schwarzschild BH & $\DD{1}{2}$ \\
  Kerr BH & 1 \\
\end{tabular}
\symbolfootnote[0]{\tiny ${}^*$Hypothetical models}
\caption{\small Different astrophysical objects and their corresponding compactness parameter $\mathcal{C}$. }  \label{Table1}
\end{centering}
\end{table}

\begin{enumerate}
  \item    While the gravastar model is based on a fluid approach, the connection to quantum effects in vacuum
   has been performed through the integration constant $\epsilon$ solely, which is rather an ad hoc approach. It is unclear,
   what are those microscopic quantum effects that would lead to different integration constants for
   different masses in a universe of well-defined universal constants.
   A reasonable analysis should ensure a scaling out of the dependence of $\epsilon$ on the mass in
   order to yield a universal constant that applies for a reasonable range of the mass function  characterizing
   astrophysical BHs.

   \item Similarly is the proposed formation of gravitational Bose-Einstein condensates -GBECs and superfluidity of
    their cores.
   Although globally stable condensates have not been verified experimentally yet, these are expected to
   form when the constituents are cooled down to a temperature near absolute zero. The particles then
   congregate into a single macroscopic quantum state. In this picture, the
   superfluid condensate is analogous to a vacuum state, its excited states to normal matter
   and its surface to a quantum critical shell. Following \citep{Chapline2004}, when ordinary
   particles enter the quantum critical shell they morph into heavy Bosonic particles.

   In fact, the long time scale of the post-glitch recovery of the Crab and Vela pulsars is a strong
    observational evidence that superfluidity might be a natural phase governing the flow dynamics
    in the cores of relativistic neutron stars.  Although the core's temperature in a NS is of order several million degrees,
    this is still two to three orders of magnitude lower than the dominant electron Fermi or the effective Coulomb temperatures
    characterizing  the core's matter. These conditions are equivalent to a terrestrial superfluid with $T \approx 10^{-3}\,K.$
    However, superfluidity in cores of pulsars is a favored phase due to presence of neutron-proton
    two-fluid nature with $n_n/n_p \approx 30,$ which gives rise to n-n and p-p pairing. Given the high rotational speed
    of pulsars, the superfluid ought to consist of discrete array of quantized vortex lines elongated parallel to their
    rotation axis. The Crab pulsar, for example, is expected to contain $N\approx 5.3\times10^{18}$ vortex lines.
    Due to the superconductivity of the core's matter, those vortex lines that are coupled to the magnetic field must
    migrate randomly outwards, leading to the loss of radiation reaction torques and subsequently decelerating the
    core's rotational speed \citep{Alpar1988}.\\

    When extending this scenario to SMBECs several difficulties may emerge, that could potentially prohibit their formation.

    Consider for example the time-evolution of the pressure, P(t),  at the center of a neutron star.
    While the dominant contribution is due to the non-thermal degenerate pressure, these objects have still to spend several
    million years in order to liberate the vast thermal energy trapped in the core during the dynamical collapse of the progenitor.
   A newly born neutron star  is expected to have a central temperature of $T_C \approx 10^{11}- 10^{12}$ K, but which decreases
   quickly thereafter to reach several million degrees  through extensive neutrino emission. Without extensive neutrino emission,
   the stored thermal energy in the core would alter the force balance and may lead to completely different evolutionary tracks.
   In particular, the superfluidity would diminish.

{As the progenitors of a stellar mass Bose-Einstein condensate (BECs) must be much more massive than that of a NS, the total thermal
   energy trapped in the core is of order $\mathcal{U} \approx V^2_s\, M_{BEC},$ where $V_s$ is the
   sound speed, assuming isothermal core. As this thermal energy would neither disappear in a singularity, as the case in BHs,
   nor being expelled to the surrounding regions\footnote{Due to the extremely large gravitational redshift}, $\mathcal{U}$ must eventually be
    comparable to the rest energy of the core. Thus $V_s$ is relativistic and therefore is much larger than  the
     critical speed, beyond which superfluidity will be destroyed. Furthermore, as the crust is made of normal
     matter, it would serve as source for electromagnetic radiation. This loss of energy would cause the core to shrink and
     subsequently collapse into a BH.\\
     We note that in the absence of self-gravity,  it was experimentally
   verified that external MFs tend to shrink the condensates,  enhance the self-interactions and reconnection of vortices
   and subsequently lead to their collapse or explosion as bosonova \citep{Wieman2001}.}

   {In the case that the BEC-phase is reached via a quasi-stationary contraction of a normal matter core,
   then the heat capacity of the matter in the crust must have been continuously decreasing to reach a small, but still
   a positive critical value, below   which the BEC would  cease to thermally interact with the surrounding media. Only a negligibly small fraction of
     $\xi = \sqrt{1- 2\mathcal{C}}$ of the total internal energy would find its way outwards, while the rest is being trapped
     in the core or diffuses backwards from the crust into the core.}

    {To conclude: the enormous thermal energy trapped in the core during the collapse of the progenitor in combination with magnetic
    fields and large conversion efficiency of  kinetic into thermal energy via shocks at the surface would eventually  act to suppress
     the superfluidity and superconductivity of the core and lead to a reverse phase transition.  }

    A further uncertainty of SMBECs is the causality problem. Consider for example the SMBH powering the high red-shifted quasar
    PKS 1020-103, whose mass estimated to be $M \approx 2.62\times 10^9 M_\odot$ and accretion rate of several solar masses per year.
    The light crossing time of this object is $\tau_{LCT} \sim 2\,r_s/c \approx \text{7}~ hours$,
    which is the shortest possible time scale required for a global re-adjustment of the core to external perturbations.\\

    Let us consider the case, in  which the core consists of a single bosonic condensate with a specific macroscopic quantum state.
      Unlike white dwarfs and neutron stars, in which the Pauli exclusion principle prevent their collapse,
      in the case of boson-like condensates it is the Heisenberg uncertainty principle that apposes self-collapse.
      The zero-energy state would allow the density of bosonic objects to be much larger than in fermionic objects.
      The critical mass of bosonic objects most likely obeys the correlation: $M_{CB} \sim M^2_{P}/m_B, $ and
      $M_{CF} \sim M^2_{P}/m^2_F$ for fermionic objects. $m_B$ and $m_F$ here denote the mass of an individual bosonic
      and fermionic particles, respectively. Provided that $m_B$ is much less than the proton mass, boson objects in principle could be
      much more massive than their fermionic counterparts.

      If a massive boson condensate consists of one single vortex line with radius $r_s$ and rotational frequency $\Omega_{BEC},$ then
      it is easy to verify from the integral of the quantized circulation:

       \[\oint V\cdot d l = \DD{h}{2 m} n, \]
      \beq
       \text{that:~~~~~~~ } \Omega_{BEC}= 6\times 10^{-72}\,(\DD{M_\odot}{M_{BEC}})^3.
      \eeq
      Here "V, l, h, m, n" correspond to the circulation velocity, path around the vortex, Planck constant, the mass involved
      in the vortex core and the number of vortices, respectively.\\

      Unless SMBEC are born missive, their cosmological growth will normally be associated with an increase of the rotational energy,
      which enforces the cores to rotate with a speed  much higher  than $\Omega_{BEC}.$

      Let us now consider the case, in which the core of the above-mentioned quasar is being hit by a neutron star.
      Noting that the mean surface density of the quasar is: $<\Sigma>_{BEC} = M/(4 \pi\, r^2_s) \approx 10^{11}\, g\,cm^{-2},$
      then the ratio of the surface energy density of the NS to that of the SMBEC at the time of the crash is:
      \beq
            \rm \DD{E^\Sigma_{NS}}{E^\Sigma_{BEC}} \approx 10^9.
      \eeq

\begin{figure}
\centering {\hspace*{-0.2cm}
\includegraphics*[angle=-0, width=5.50cm, bb=0 0 365 168,clip]{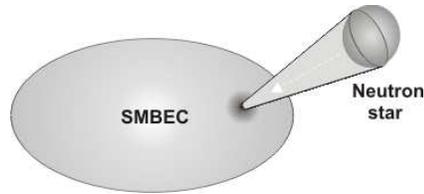}}
\caption{\small A frontal crash of a neutron star with a non-rotating supermassive
Bose-Einstein condensate.}
  \label{Fig4}
\end{figure}

       In evaluating the last expression  we have assumed $M_{NS}= M_\odot$ and $R_{NS}=10^6\,cm.$  On the other hand,
       for a gravastar, for example,
       the total mass of the membrane is roughly equal to the Planck mass. Thus the mass of the portion of the membrane possibly involved directly in this crash can
       be estimated to be: $\sim M_{PL}\,(S_{NS}/S_{SMBEC}) \approx 10^{-21}\,g,$ where $S_\Box$ means the surface of the corresponding object.
       Thus, the local energy available for resisting the NS-SMBEC
       crash is of the order of several ergs. Also, the local time scale characterizing the crash event is $\tau_{LOC} \sim R_{NS}/c\approx 10^{-9}\tau_{GD},$
       where $\tau_{GD}$ is the global dynamical time scale of the SMBEC. \\
       Given that the input of energy through the crash exceeds the locally available one by several
       orders of magnitude and that the NS would crash into membrane almost with the speed of light, we conclude that
       a considerable portion of the membrane and the enclosed portion of the SMBEC would be completely destroyed early enough before the
       condensate could react dynamically to maintain global stability. In the case of stellar mass BECs, the input of energy
       associated with such a crash would destroy the whole condensate.

       \item  When feeding a gravitational BEC with the mass rate: $\dot{M} = c^3/2G,$ then the horizon would grow with  the speed of light.
       This is equivalent to inject the core with one solar mass per $10^{-5}$ seconds, which is much  longer than the duration
       of a NS-SBEC crash-event, which is expected to be of order $\epsilon\, (r_s/c)$.
       Let us assume that the membrane of a BEC be located an epsilon small outside the event horizon, i.e. at
       $r_{BEC}= (1 + \epsilon)\,r_s$  and that this configuration is applicable  to DEOs and independent of their mass, then
       it is easy to verify from the mass-radius relation that:
       \beq
       \DD{dr}{dt}|_{BEC} = \DD{2G}{c^2}(1+\epsilon)\DD{dM}{dt}> \DD{dr}{dt}|_{BH}.
       \eeq

       Consequently, in order to ensure that the surface of the newly formed core still lays outside the event horizon,
       its surface must contradictory grow with a superluminal speed.

     \item The settling normal matter from the surrounding would create a multi-component fluid in the crust, establishing herewith the
           appropriate conditions for the creation of different Fermi surfaces for fermions and bosons and therefore applying a
           magnetic tension, which would tend to couple the core to the crust dynamically.
           As a consequence,  the core break into a multiple number of vortex lines threaded by magnetic fields.
           Similar to the superfluid and superconducting cores in NSs, the vortex lines in the core of a SMBEC that are
           coupled to the magnetic field would migrate from inside-to-outside, enhancing thereby  self-interaction
           and vortex reconnection and diminishing thereby the  superfluidity of the core.
            In this case, we anticipate the surface of a SMBEC to be magnetically active with intense eruption events, though
           hardly observable. Similar to cores in NSs, the core-crust coupling through magnetic fields would cause
           SMBEC to decelerate its rotational speed.

            Moreover, we note that rotating superfluids in BEC-cores most likely are similar to rotating superfluids between concentric spheres.
            The latter  was verified to be unstable against non-axisymmetric perturbations compared to normal rotating Couette
            flows \citep{Barenohi1987}.

\begin{figure}
\centering {\hspace*{-0.2cm}
\includegraphics*[angle=-0, width=7.50cm, bb=25 181 570 600,clip]{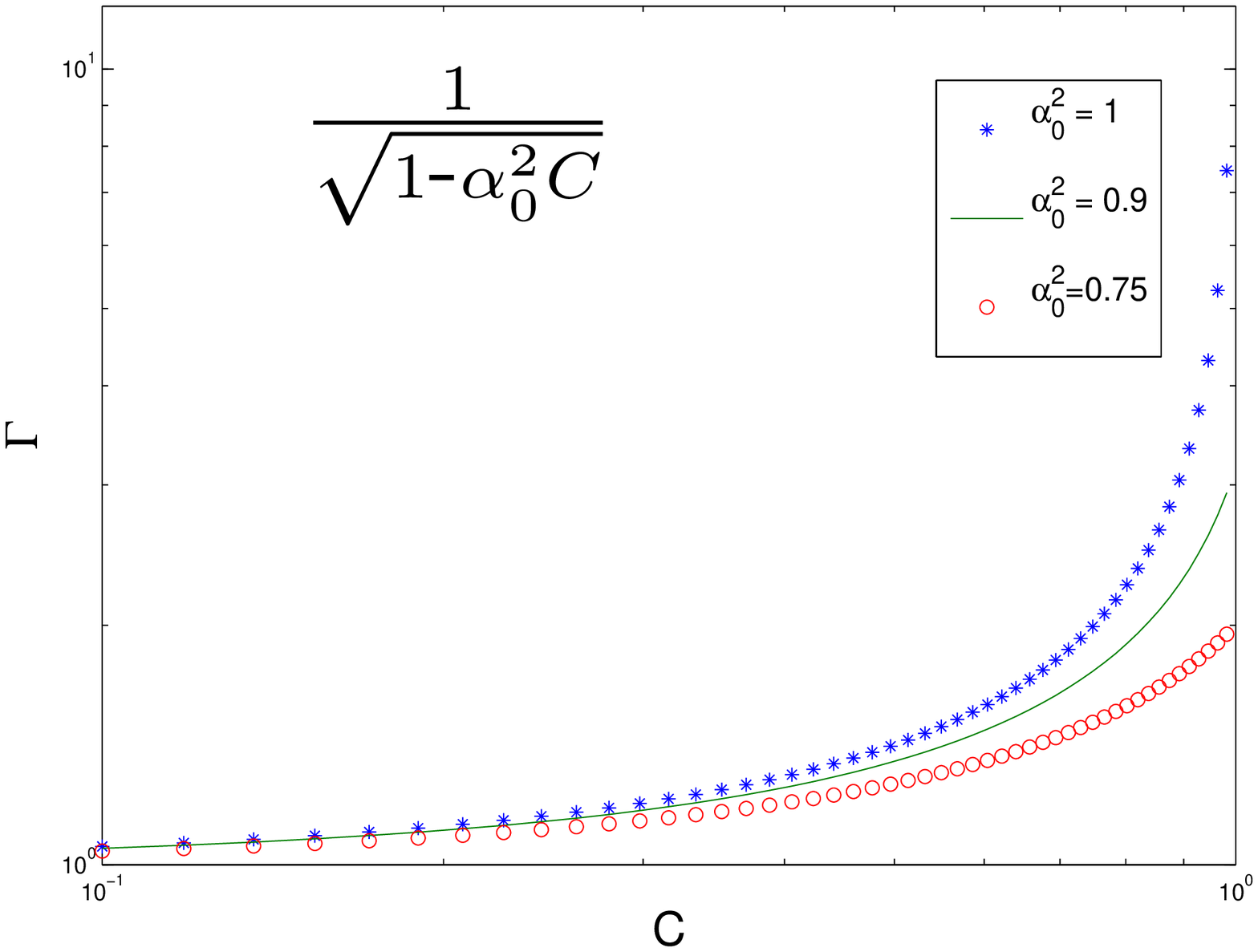}}
\caption{\small  The Lorentz factor versus the compactness parameter $\mathcal{C}$ is plotted for three different jet launching parameters: $\alpha^2_0 = 1, 0.9, 0.75$ denoted
  with the black, blue and red colors, respectively.}  \label{LorentzFact}
\end{figure}

\item  Highly collimated relativistic jets have been observed to emanate from the vicinity of accreting BHs and NSs. The Lorentz factors
       characterizing the propagational speed of these jets were verified to correlate with the compactness parameter of  accreting
        objects \citep[see][and the references therein]{hujeirat2002, hujeirat2003}. To first order approximation, we may assume jet-velocities to linearly correlate with the escape velocity, i.e.,  $V_J = \alpha_0\,V_{es},$ whereat $\alpha_0$ is a constant coefficient of order unity.
        The Lorentz factor  can be expressed then as function of the compactness parameter as follows: $\Gamma = 1/\sqrt(1-\alpha^2_0 \mathcal{C}).$ Figure (\ref{LorentzFact}) shows that $\Gamma$  is highly sensitive to $\alpha_0,$ indicating therefore that large $\Gamma-$factors between 10 to 20 may be obtained if the jet velocity is a significant fraction of the escape velocity. Equivalently, the jet-plasma must star its
        outwards-oriented motion from the very vicinity of the object's surface or from nearly the event horizon.  \\

        We note that the formation and acceleration of jet-plasmas around accreting relativistic objects is considered to be the outcome of
        complicated coupling processes between the accretion disk, the central object and the jet-plasma, whereby magnetic fields
        play a crucial role \citep{hujeirat2004, hujeirat2011}.
        Jets emanating from around surface-free objects are generally observed to be radio-dominated and their plasmas
        propagate with larger Lorentz factors compared to those of neutron stars. Therefore, in the case of accreting DEOs the solid surface
        in combination with magnetic fields threading the normal-matter-made membrane would give rise to an equally significant
        thermal and radio emission with strong variabilities. These could serve as guide to fix the compactness parameter of the object.
        However, a stellar type DEO would display variabilities whose maximum frequency behaves as:
        \[ \nu_{QPO} \approx \DD{\sqrt{\mathcal{C}}}{2\pi} \DD{c}{r}.\]
        If the microquasar GRS 1915+ 105 were a DEO, for example,
         then it would display quasi-periodic oscillation around 400 Hz. This is approximately one order of magnitude larger than the
         revealed value by observations. Moreover, the dominant radio over thermal emission characterizing this objects rules out the
         possibility of a solid surface.

\item  The intensity of the magnetic fields must amplify to reach roughly equipartition values as the accreted plasma
        shocks the crust of the SMBEC (Fig. \ref{BEC}). Consequently, a boundary layer must form, whose thickness, $\ell,$ scales as:
        \[ \DD{\ell}{r_\star} \sim \mathcal{C} (\DD{V_A}{c})^2.\]
        $V_A$ in this correlation stands for the  Alf$\grave{v}$en speed. Thus,  the boundary layer (BL), which is most
        likely filled with virially hot elementary particles, is effectively a photosphere with a macroscopic thickness
        rather than a membrane with sub-atomic microscopic width. In this case, large scale magnetic fields are capable of
        communicating the presence of the solid crust to the surrounding media. The freely falling charged particles would then decelerate
        in the BL, emitting thereby a considerable amount of their energy at the synchrotron frequency and give rise
        to a total synchrotron luminosity of the order:
        \[L_{syn} \sim 10^{45}~ n^2~ (\DD{M_9}{T_4})^3 \,erg\, s^{-1}.\] $n,\, T_4,\, M_9$ correspond to the number density, the
        pre-shock temperature of the accreted plasma and the mass of the SMBEC in units of $10^9\,M_\odot,$ respectively.\\
        Consequently, such a large radio luminosity from a geometrically thin BL would be optically thick to
        synchrotron emission and the BL would appear as a bright photospheric ring that surrounds the SMBEC.
        However, such rings have not been observed yet, though they would be comfortably observable with today's
        detectors.

         \end{enumerate}

\begin{figure}
\centering {\hspace*{-0.2cm}
\includegraphics*[angle=-0, width=6.85cm, bb=0 0 424 162,clip]{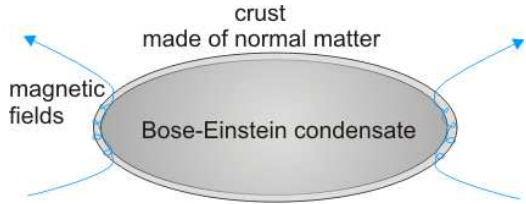}}
\caption{\small  Magnetic fields threading the matter made crust (; unscaled thickness) of the condensate.}  \label{BEC}
\end{figure}

\section{Summary}

Various aspects of black holes and dark energy objects as BH-candidates have been discussed
and the draw backs of both kind of objects have been addressed.  In particular, it is argued that
gravitational Bose-Einstein condensates, as alternatives to supermassive black holes,
suffers of a causal problem and should be ruled out on the case of merger events.
Moreover, the superfluidity of SMBEC-cores most likely is a short-living phase and would
not survive non-axisymmetric perturbations initiated by the magnetic fields, external forces or mergers.
Such perturbations would enhance the vortex-line migration phenomena, increase their rate of interaction and reconnection
 and finally turn the core dissipative, leading subsequently to core collapse into a black hole or explode as bosonova.

\vspace*{0.5cm}

\noindent{\bf Acknowledgments}  Sofie Felhmann (University of Basel) is greatly acknowledged for
reproducing part of the Figures. \\

\clearpage

\renewcommand{\theequation}{\thesection{}\arabic{equation}}

\end{document}